\documentclass[reprint,superscriptaddress,preprintnumbers,
groupedaddress,
nofootinbib,
 amsmath,amssymb,
 aps]{revtex4-1}
\usepackage{multirow}
\usepackage{upgreek}
\usepackage{tensor}
\usepackage{float}
\usepackage{latexsym}
\usepackage{graphicx}
\usepackage{amssymb}
\usepackage{amsmath}
\usepackage{amsfonts}
\usepackage[hidelinks]{hyperref}
\usepackage{color}
\usepackage{cool}
\usepackage{dcolumn}
\usepackage{slashed}
\usepackage{mathrsfs}
\usepackage{subcaption}

\begin{document}

\title{A note on interacting holographic dark energy with a Hubble-scale cutoff}


\author{Ricardo G. Landim}\email{ricardo.landim@tum.de}

\affiliation{Technische Universit\"at M\"unchen, Physik-Department T70,\\ James-Franck-Stra\text{$\beta$}e 1, 85748 Garching, Germany}


\date{\today}

\begin{abstract}
Holographic dark energy with the Hubble radius as infrared cutoff has been considered as a candidate to explain the late-time cosmic acceleration and it can solve the coincidence problem. In this scenario, a non-zero equation of state is only possible if there is an interaction between dark energy and cold dark matter. In this paper, a set of phenomenological interactions is assumed and a detailed analysis of the possible values of the coupling constants is carried out, however the resulting matter power spectrum and cosmic microwave background temperature and polarization power spectra have a shape very far from the observed ones. These results  rule out any value for the free parameters and it seems to indicate that the assumed interacting holographic dark energy with a Hubble-scale cutoff is not  viable to explain the accelerated expansion of the Universe, when cosmological data are taken into account. 
  \end{abstract}

\maketitle

\section{Introduction}

The observational evidence of dark energy (DE) in 1998 \cite{perlmutter1999,reiss1998} opened a new phase in the understanding of our Universe. While cosmological data are continuously confirming the existence of the late-time cosmic acceleration (see \cite{copeland2006dynamics,pdg2022} for reviews), the nature of the accelerated expansion is still an open issue.  The simplest candidate for DE is a cosmological constant $\Lambda$, which encompasses the standard $\Lambda$-cold-dark-matter (CDM) model. The six free parameters of the $\Lambda$CDM model are well constrained and are in agreement with cosmological observations \cite{Aghanim:2018eyx}, despite some tensions, e.g. the Hubble tension \cite{Aghanim:2018eyx,Riess:2021jrx} (a recent review is \cite{DiValentino:2021izs}). The  observed value of the vacuum energy is many orders of magnitude smaller than the  theoretically calculated \cite{Weinberg:1988cp}, leading to the so-called `cosmological constant problem'. Additionally, the evolution of CDM and DE are very different from each other, but their energy densities today have the same order of magnitude. This coincidence may indicate new physics and it is usually referred to as `coincidence problem'. 

The lack of understating about the nature of the cosmological constant and the aforementioned issues encourage alternative models of DE (for reviews see \cite{copeland2006dynamics,Bamba:2012cp}). Among the many candidates there are   scalar and vector fields \cite{peebles1988,ratra1988,Frieman1992,Frieman1995,Caldwell:1997ii,Padmanabhan:2002cp,Bagla:2002yn,ArmendarizPicon:2000dh,Brax1999,Copeland2000,Vagnozzi:2018jhn,Koivisto:2008xf,Bamba:2008ja,Emelyanov:2011ze,Emelyanov:2011wn,Emelyanov:2011kn,Kouwn:2015cdw,Landim:2015upa,Landim:2016dxh,Banerjee:2020xcn}, metastable DE \cite{Szydlowski:2017wlv,Stachowski:2016zpq,Stojkovic:2007dw,Greenwood:2008qp,Abdalla:2012ug,Shafieloo:2016bpk,Landim:2016isc, Landim:2017kyz,Landim:2017lyq}, models using extra dimensions \cite{dvali2000}, alternative fluids \cite{Landim:2021www, Landim:2021ial}, etc. Another explanation for DE comes from the holographic principle, the so-called holographic DE (HDE) \cite{Hsu:2004ri,Li:2004rb,Pavon:2005yx,Nojiri:2005pu,Wang:2005jx,Wang:2005pk,Wang:2005ph,Wang:2007ak,Landim:2015hqa,Li:2009bn,Li:2009zs,Li:2011sd,Saridakis:2017rdo,Mamon:2017crm,Mukherjee:2016lor,Feng:2016djj,Herrera:2016uci,Forte:2016ben,DAgostino:2019wko,Nojiri:2020wmh,Nojiri:2021iko,Nojiri:2021jxf, Colgain:2021beg} (see \cite{Wang:2016och} for a review). The holographic principle suggested by 't Hooft \cite{'tHooft:1993gx} and Susskind \cite{Susskind:1993aa, Susskind:1994vu}, in turn based on the previous works of Thorn \cite{Thorn:1991fv} and Bekenstein \cite{Bekenstein:1993dz},  is a property of quantum gravity, where at Planckian scale the  world is best described by a 2-D lattice evolving with time, rather than 3+1-D. In this scenario, DE should obey this principle  and the fine-tuning problem is eliminated \cite{Cohen:1998zx}. 

HDE would then have an energy density given by $\rho_{\rm de}=  3c^2M_{Pl}^2L^{-2}$, where $c$ is a constant, $M_{Pl}$ is the reduced Planck mass and $L$ is the infrared (IR) cutoff \cite{Hsu:2004ri, Li:2004rb}. The first natural choice for the IR cutoff is the Hubble radius, however it led to an equation of state that described pressureless matter \cite{Hsu:2004ri}. This problem was circumvented choosing the future event horizon as cutoff \cite{Li:2004rb}. Other choices for $L$ include the inverse of the Ricci scalar curvature \cite{Gao:2007ep}, the age of the Universe \cite{Cai:2007us}, among others \cite{Saridakis:2020zol,Nojiri:2022aof,Nojiri:2022dkr}. Inspired by the holographic principle and the AdS/CFT correspondence \cite{Maldacena:1997re}, HDE has been embedded in minimal supergravity \cite{Landim:2015hqa}, while in \cite{Nastase:2016sji} it was shown that HDE arises from generic quantum gravity theory, assuming only the existence of a minimum length.

Another widely studied alternative to the $\Lambda$CDM paradigm is if DE interacts with CDM \cite{Wetterich:1994bg,Amendola:1999er,Guo:2004vg,Cai:2004dk,Guo:2004xx,Bi:2004ns,Gumjudpai:2005ry,Sadjadi:2006qb,Yin:2007vq,MohseniSadjadi:2008na,Costa:2013sva,Abdalla:2014cla,Costa:2014pba,Landim:2015poa,Landim:2015uda,Costa:2016tpb,Marcondes:2016reb,Landim:2016gpz,Wang:2016lxa,Farrar:2003uw,micheletti2009,Yang:2017yme,Marttens:2016cba,Yang:2017zjs,Costa:2018aoy,Yang:2018euj,Landim:2019lvl,Vagnozzi:2019kvw} and it can help alleviating the coincidence problem \cite{Olivares:2005tb} and the Hubble tension \cite{DiValentino:2019ffd,DiValentino:2019jae,Lucca:2020zjb}. Among the many  possible phenomenological interactions, one of the most famous and used in the literature is at the background level, proportional to the sum of the energy densities of CDM and DE (see \cite{Wang:2016lxa} for a review). Constraints on the couplings constants and forecasts  for several upcoming observational programs are presented in \cite{Costa:2013sva, Costa:2016tpb, Costa:2019uvk, Bachega:2019fki, Costa:2021jsk,Xiao:2021nmk, Landim:2022}.

Assuming an interaction between HDE (with a Hubble radius as IR cutoff) and CDM not only gives the correct equation of state for DE, but also solves the coincidence problem \cite{Pavon:2005yx}. In this paper we investigate  this HDE model, using the aforementioned phenomenological interactions. We perform a detailed analysis of  the necessary values for the couplings that would give an equation of state in agreement with the cosmic acceleration. It turns out that the parameter space leads to a matter and cosmic microwave background (CMB)  power spectra in disagreement to what is observed. The resulting power spectra are actually very similar to the $\Lambda$CDM model but without CDM, therefore not being able to reproduce current cosmological observations.

This paper is organized as follows. Sec. \ref{sec:idedm} reviews some aspects of the HDE model considered here, along with the phenomenological interactions, and present the necessary equations. In Sec. \ref{sec:results} we show   our   results and Sec. \ref{sec:conclusions} is reserved for conclusions. We use Natural units ($c=\hbar=1$) throughout the text.

\section{Holographic dark energy}\label{sec:idedm}
When the Hubble scale is considered as IR cutoff $L^{-1}=H$, the energy density for DE is $\rho_{\rm de}=  3c^2M_{Pl}^2H^2$, while for CDM the energy density becomes $\rho_{\rm dm} = 3(1-c^2)M_{Pl}^2H^2$, where the first Friedmann  equation for a spatially flat Universe was used, safely ignoring radiation and visible matter. The ratio $r\equiv \rho_{\rm dm}/\rho_{\rm de}$ is thus $r= (1-c^2)/c^2$ \cite{Pavon:2005yx}, therefore constant if $c$ is constant. When there is an interaction between DE and CDM the total energy momentum tensor is still conserved, however not anymore for the individual components. The continuity equations are
\begin{align}
    \dot{\rho}_{\rm dm}+3H\rho_{\rm dm}&= Q\,,\\
     \dot{\rho}_{\rm de}+3H(1+w)\rho_{\rm de}&= - Q\label{eq:cont_de}\,,
\end{align}
where $w$ is the constant DE equation of state and a dot represents a time derivative. We take the phenomenological interaction   $Q=H(\lambda_1\rho_{\rm dm}+\lambda_2\rho_{\rm de})$ \cite{He:2008si}, where $\lambda_1$ and $\lambda_2$ are constants. 

Here we will use the original scenario of constant $c$. Using the expression for $\rho_{\rm de}$ into Eq. (\ref{eq:cont_de}) the equation of state is determined
\begin{equation}\label{eq:w_ihde}
    w=-\frac{1}{3}\Big(\lambda_1+\frac{\lambda_2}{r}\Big)(1+r)\,.
\end{equation}
This means that the equation of state is no longer a free parameter, as it is usual in other interacting DE models. When the coupling constants are zero, a pressureless fluid is recovered, as originally found in \cite{Hsu:2004ri}.  

The equation of state is constant and depends on the coupling constants, given that the ratio $r$ is well known. In order for $w$ not to be zero, the coupling constants should not be very small. The constant $c$ is also completely determined by the ratio $r=r_0$, through $c^2=(1+r_0)^{-1}$.
 
 We can solve the corresponding continuity equations, which give the energy densities for CDM and DE, respectively,
 \begin{align}
\rho_{\rm dm} &=  \rho_{\rm dm, 0}a^{-3   +\lambda_1+ \frac{\lambda_2}{r_0}}\,,\\
\rho_{\rm de} & = \rho_{\rm de,0}a^{-3 +\lambda_1+ \frac{\lambda_2}{r_0}}\,.
 \end{align}
  Both CDM and DE present the same background evolution, with an effective equation of state $w_{\rm de}^{\rm eff}=w_{\rm dm}^{\rm eff}= -1/3(\lambda_1+\lambda_2/r_0)$, thus leading to  the constant ratio $r$ at all times. This already poses a problem in the description, because both fluids will have the same evolution, therefore they both describe either CDM (with $\lambda_1=\lambda_2\simeq0$) or DE ($w^{\rm eff}_{\rm de}<-1/3$). The  first scenario is what Hsu found \cite{Hsu:2004ri} and the second one is the incentive to add an interaction in the first place. However, if both fluids describe DE then we would have a Universe without CDM, which is ruled out by observations. We will return to this point in a moment.
  
\begin{figure}[b]
      \centering
      \includegraphics[scale=0.55]{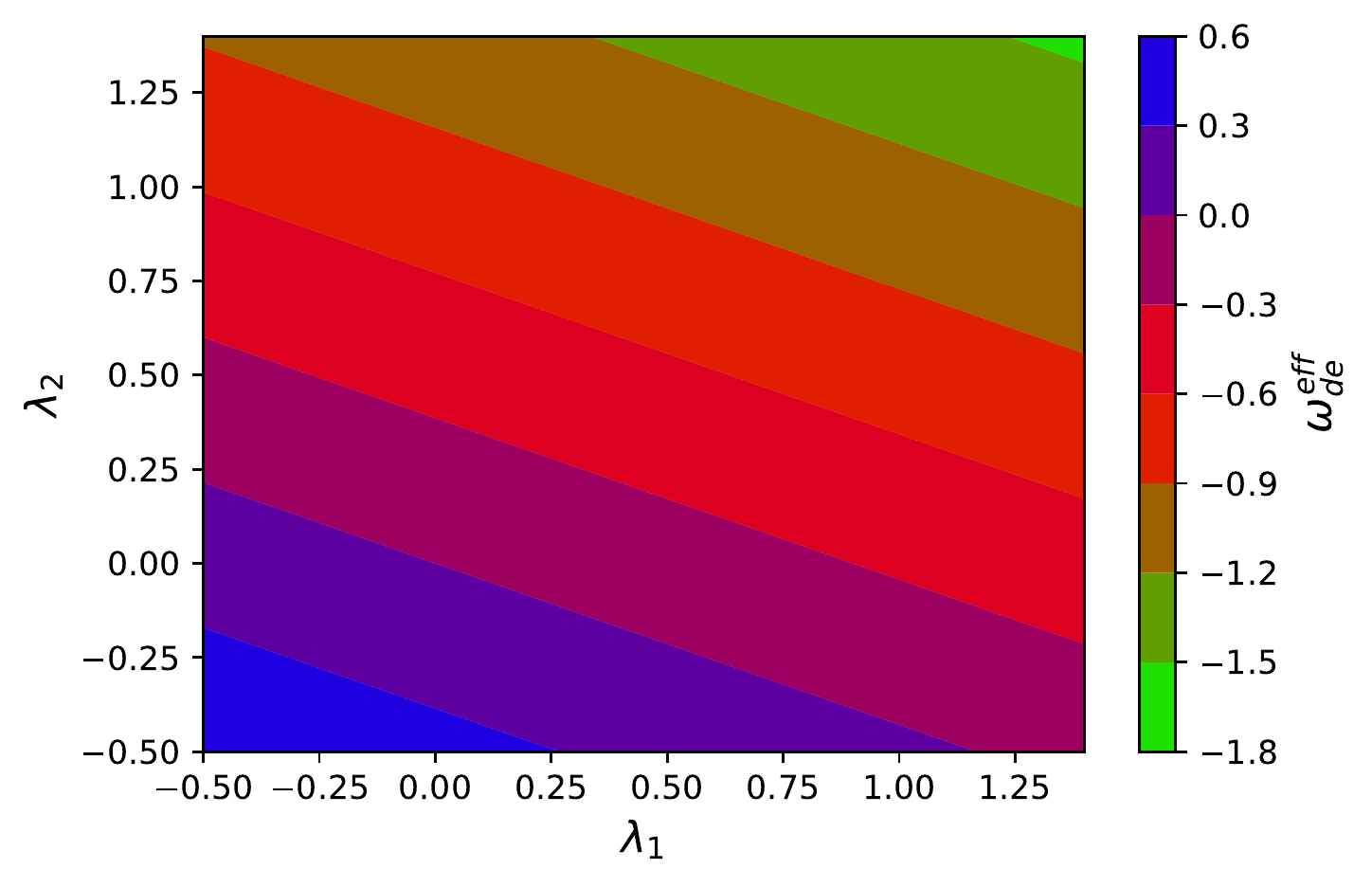}
      \caption{Effective equation of state for DE as a function of the coupling constants. The accelerated expansion of the Universe may happen for relatively large couplings. If $\lambda_1 = 0$, then $\lambda_2$ should be considerably larger than 1.}
      \label{fig:wihde}
  \end{figure}

  The accelerated expansion can only be achieved if both couplings are not very small, as it is depicted in Fig. \ref{fig:wihde}.  One may wonder if such relatively large couplings are in agreement with observations, since in other models the couplings are small \cite{Costa:2016tpb}. This issue is investigated as follows.

Although HDE is an effective description for the cosmological constant, the perturbation of the energy density is non-zero, in contrast to  $\Lambda$CDM. The perturbation is $\delta_{\rm de}=2\delta H/H$, where the perturbation in the Hubble rate is given by $\delta H = k v_T/3 + \dot{h}/6$ \cite{Gavela:2010tm}. We can use the full set of linear order perturbation equations for CDM and DE to investigate the CDM behavior. In the synchronous gauge they are \cite{Ma:1995ey,Valiviita:2008iv,Gavela:2010tm,Costa:2013sva,Landim:2022}
\begin{widetext}
\begin{align}
    \dot{\delta}_{\rm dm} =& -\theta_{\rm dm}-\frac{\dot{h}}{2}+\mathcal{H}\lambda_2\frac{\rho_{\rm de,0}}{\rho_{\rm dm,0}}(\delta_{\rm de}-\delta_{\rm dm})+\bigg(\lambda_1+\lambda_2\frac{\rho_{\rm de,0}}{\rho_{\rm dm,0}}\bigg)\bigg(\frac{kv_T}{3}+\frac{\dot{h}}{6}\bigg)\,,\label{eq:delta_dm}\\
    \dot{\theta}_{\rm dm} = & -\mathcal{H}\theta_{\rm dm}-\bigg(\lambda_1+\lambda_2\frac{\rho_{\rm de,0}}{\rho_{\rm dm,0}}\bigg)\mathcal{H}\theta_{\rm dm}\,,\\
    \dot{\delta}_{\rm de} =& -(1+w)\bigg(\theta_{\rm de}+\frac{\dot{h}}{2}\bigg)-3\mathcal{H}(1-w)\delta_{\rm de}+\mathcal{H}\lambda_1\frac{\rho_{\rm dm,0}}{\rho_{\rm de,0}}(\delta_{\rm de}-\delta_{\rm dm})\nonumber\\&-3\mathcal{H}(1-w)\bigg[3(1+w)+\lambda_1\frac{\rho_{\rm dm,0}}{\rho_{\rm de,0}}+\lambda_2\bigg]\frac{\mathcal{H}\theta_{\rm de}}{k^2}-\bigg(\lambda_1\frac{\rho_{\rm dm,0}}{\rho_{\rm de,0}}+\lambda_2\bigg)\bigg(\frac{kv_T}{3}+\frac{\dot{h}}{6}\bigg)\,,\label{eq:delta_de}\\
    \dot{\theta}_{\rm de} = & 2\mathcal{H}\theta_{\rm de}\bigg[1+\frac{1}{1+w}\bigg(\lambda_1\frac{\rho_{\rm dm,0}}{\rho_{\rm de,0}}+\lambda_2\bigg)\bigg]+\frac{k^2}{1+w}\delta_{\rm de}\,,
\end{align}
\end{widetext}
where  the adiabatic sound speed is assumed to be $w$, the DE effective sound speed
is one and the center of mass velocity for the total fluid $v_T$ is defined as \cite{Gavela:2010tm} 
\begin{equation}
   (1+w_T) v_T=\sum_a (1+w_a)\Omega_a v_a\,.
\end{equation}
The DE equation of state $w$ is given by Eq. (\ref{eq:w_ihde}) with constant $r$.

In the synchronous gauge, the adiabatic initial conditions for CDM and DE  are \cite{He:2008si,costa2014observational}
\begin{align}
    \delta_{\rm de}^{(i)}&= \delta_{\rm dm}^{(i)}=\frac{3}{4}\delta_r^{(i)} \bigg(1-\frac{\lambda_1}{3}-\frac{\lambda_2}{3}\frac{1}{r_0}\bigg)\,,\label{eq:delta_ini_dm}\\
    v_{\rm de}^{(i)} &= v_r^{(i)}\,,
    \end{align}
where the index `r' represents radiation. The equations for the other species remain as they are in the $\Lambda$CDM model. Finally, a comoving frame where the CDM velocity is zero is chosen in order to fix the residual freedom of the synchronous gauge.   
\section{ Results}\label{sec:results}

We implemented the background and perturbation equations in a modified version of \texttt{CLASS} \cite{blas2011cosmic,Lucca:2020zjb}.

\begin{figure}
    \centering
    \includegraphics[scale=0.5]{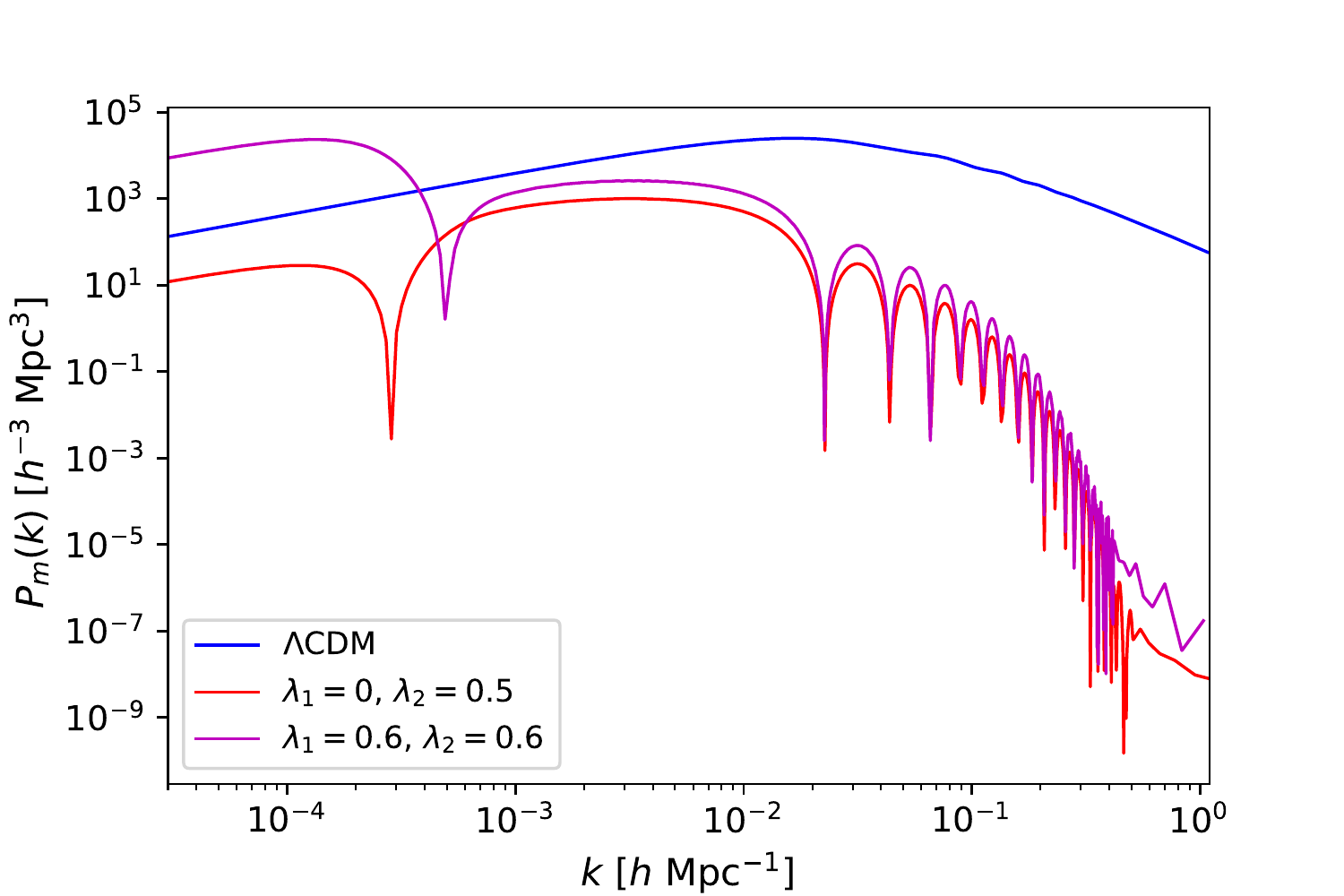}
    \caption{Linear matter power spectrum at $z=0$ for two sets of representative values for $\lambda_1$ and $\lambda_2$, and $\Lambda$CDM. The usual cosmological parameters were fixed to the \textit{Planck} 2018 best-fit values.}
    \label{fig:pk}
\end{figure}

\begin{figure*}
\centering
\begin{subfigure}{0.45\textwidth}
  \centering
  \includegraphics[width=\linewidth]{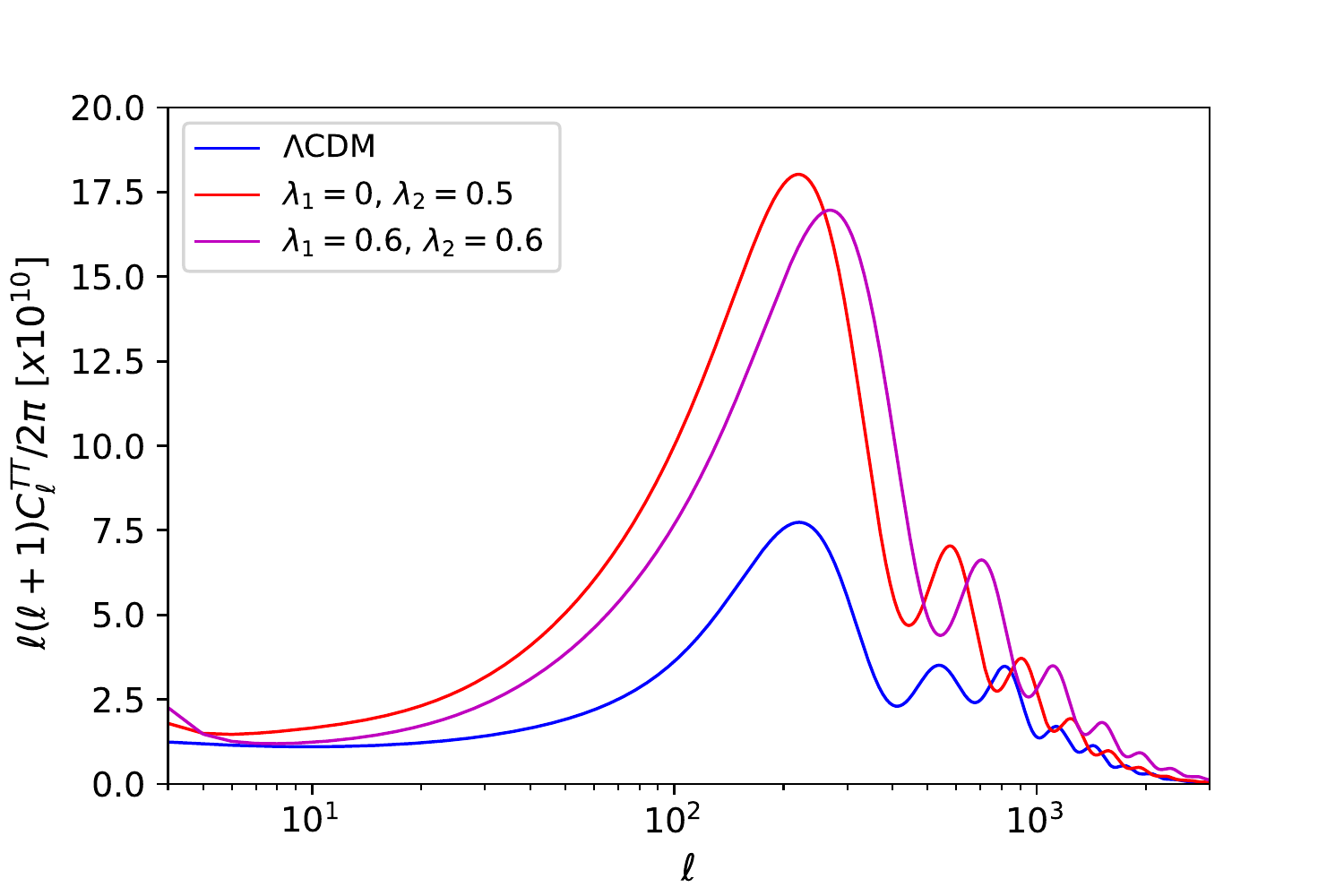}
   \label{fig:sub1}
\end{subfigure}%
\begin{subfigure}{.45\textwidth}
  \centering
  \includegraphics[width=\linewidth]{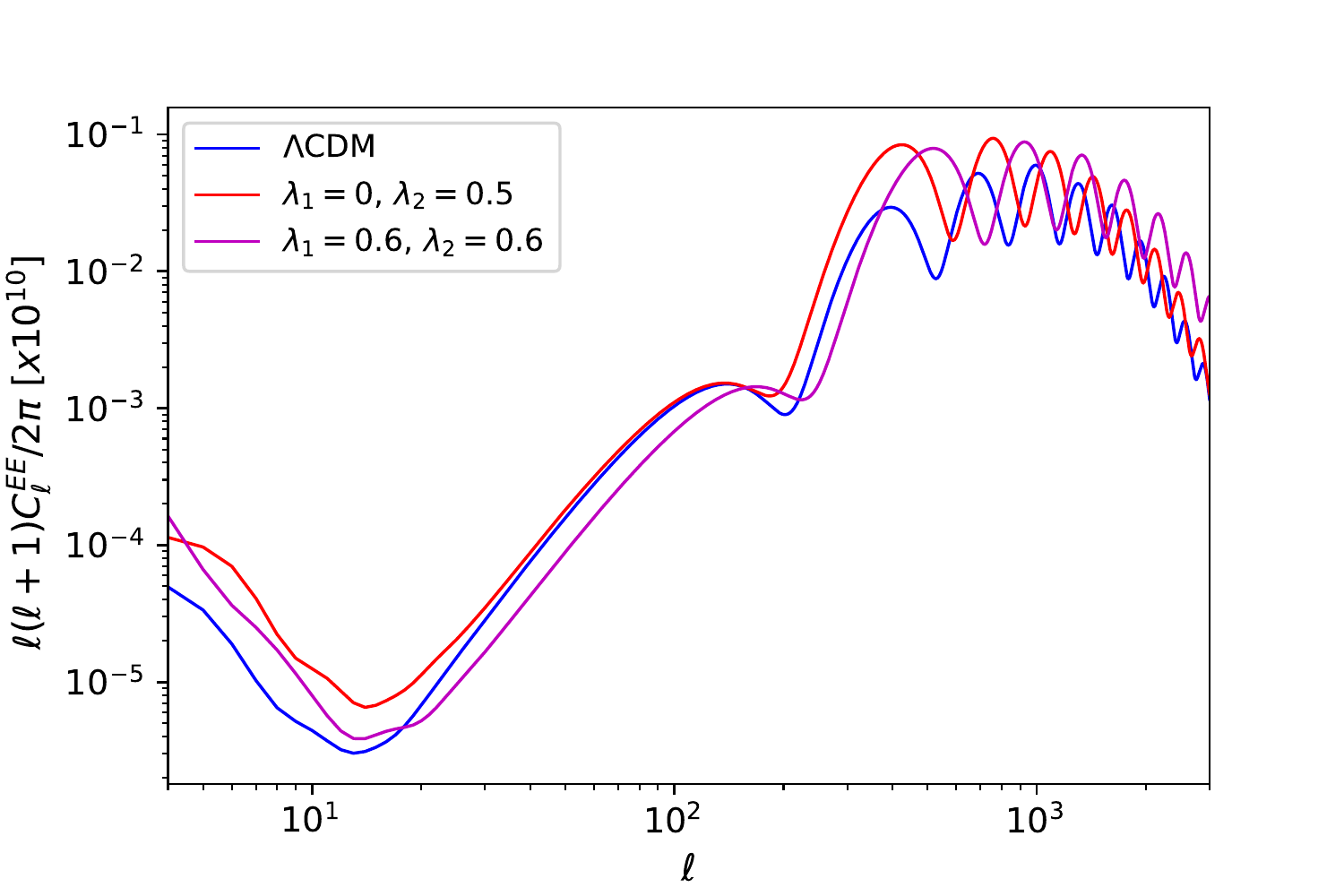}
   \label{fig:sub2}
\end{subfigure}
\begin{subfigure}{0.45\textwidth}
  \centering
  \includegraphics[width=\linewidth]{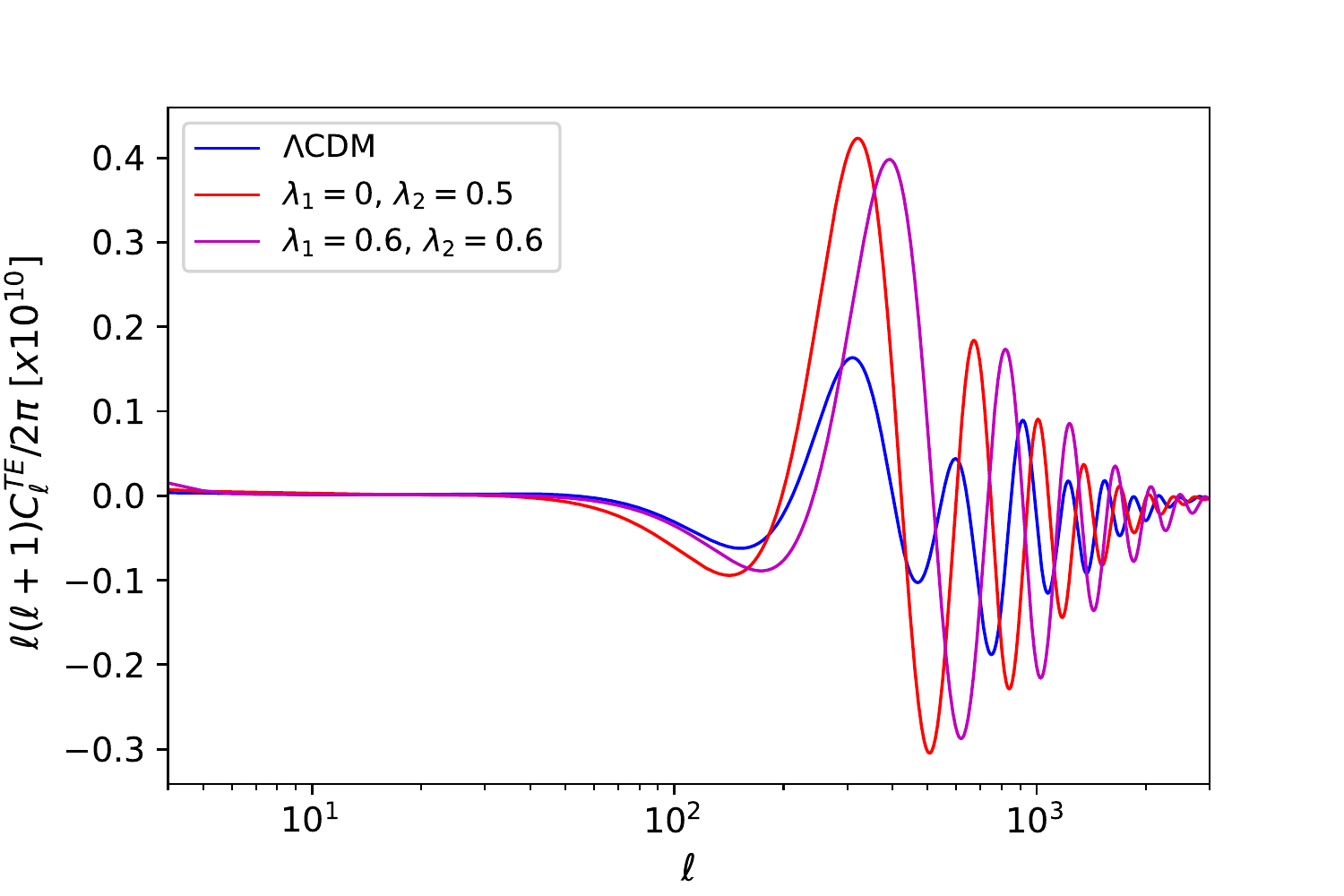}
  \end{subfigure}%
\caption{Dimensionless CMB temperature (top left), polarization (top right) and cross (bottom) power spectra. Two sets of representative values for $\lambda_1$ and $\lambda_2$ were taken, while the other cosmological parameters were fixed to the \textit{Planck} 2018 best-fit values. A comparison with $\Lambda$CDM is also shown. }
\label{fig:cl}
\end{figure*}

We have extensively investigated the parameter space, and illustrative  matter and CMB power spectra are shown in Figures \ref{fig:pk} and \ref{fig:cl}, respectively, where we chose two different set of values for $\lambda_1$ and $\lambda_2$ and plot also the case for $\Lambda$CDM. Independent of the  chosen values for the couplings, the power spectra are very different from $\Lambda$CDM.  

Several experiments have measured the CMB power spectrum since 1992, e.g. COBE \cite{COBE:1992syq}, TOCO \cite{Miller:1999qz}, DASI \cite{Halverson:2001yy}, Boomerang \cite{Boomerang:2001tye}, MAXIMA \cite{Hanany:2000qf}, WMAP \cite{WMAP:2012fli,WMAP:2012nax}, and more recently \textit{Planck} \cite{Planck:2018nkj}. All of these experiments constrained very well the CMB power spectrum, which is in agreement with the $\Lambda$CDM model. Therefore, a deviation from the observed power spectrum, like the ones shown in Fig. \ref{fig:cl}, is very disfavored. The same conclusion can be drawn for the matter power spectrum. The matter power spectrum is well constrained by latest observations, e.g. \textit{Planck} 2018 CMB data \cite{Planck:2018nkj}, DES Year 1 cosmic shear \cite{DES:2017qwj}, and SDSS galaxy and Ly $\alpha$ clustering \cite{Reid:2009xm, SDSS:2017bih,Chabanier:2018rga,Chabanier:2019eai}. Thus all choices of couplings are completely excluded from current observations. 

\begin{figure*}
\centering
\begin{subfigure}{0.45\textwidth}
  \centering
  \includegraphics[width=\linewidth]{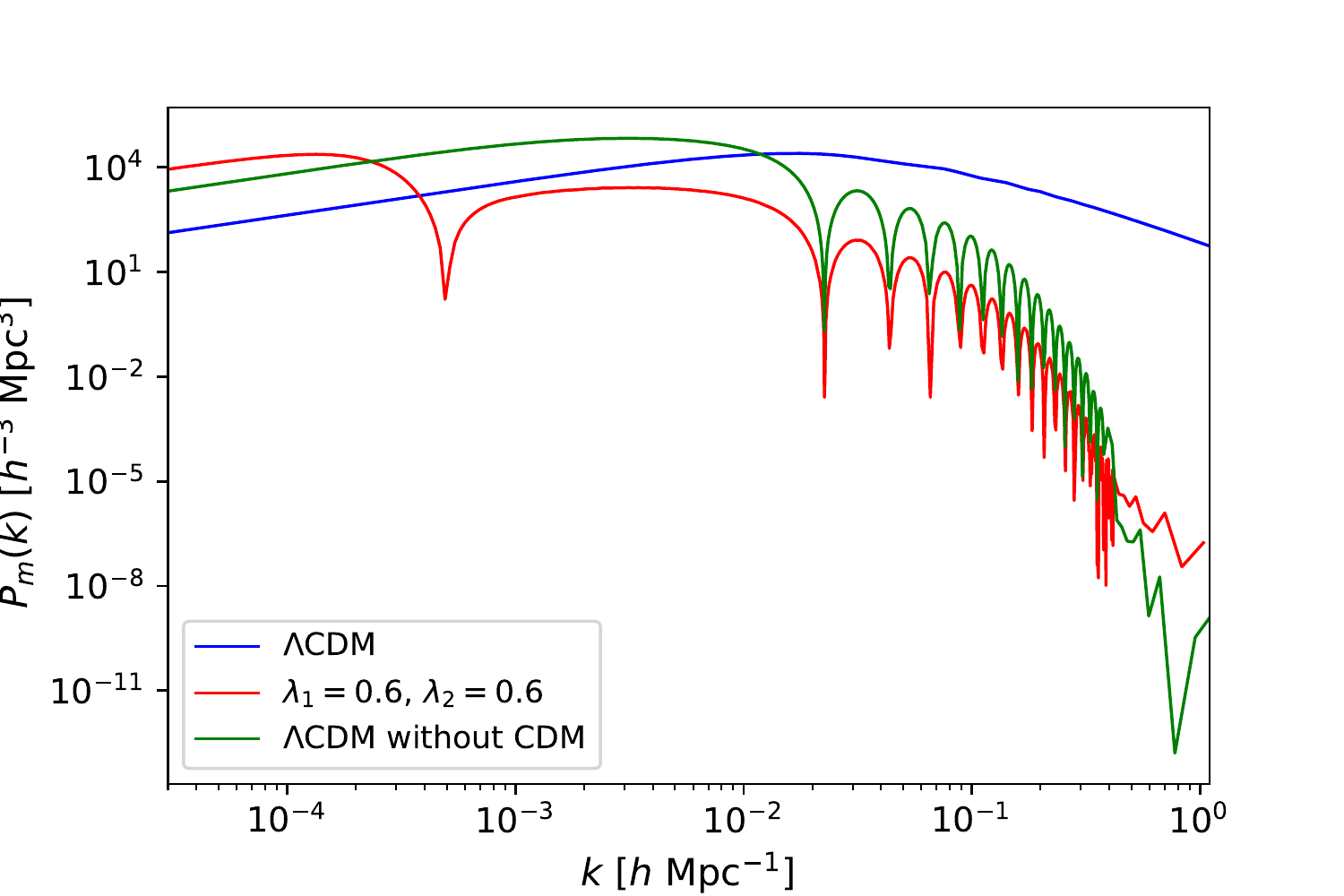}
\end{subfigure}%
\begin{subfigure}{.45\textwidth}
  \centering
  \includegraphics[width=\linewidth]{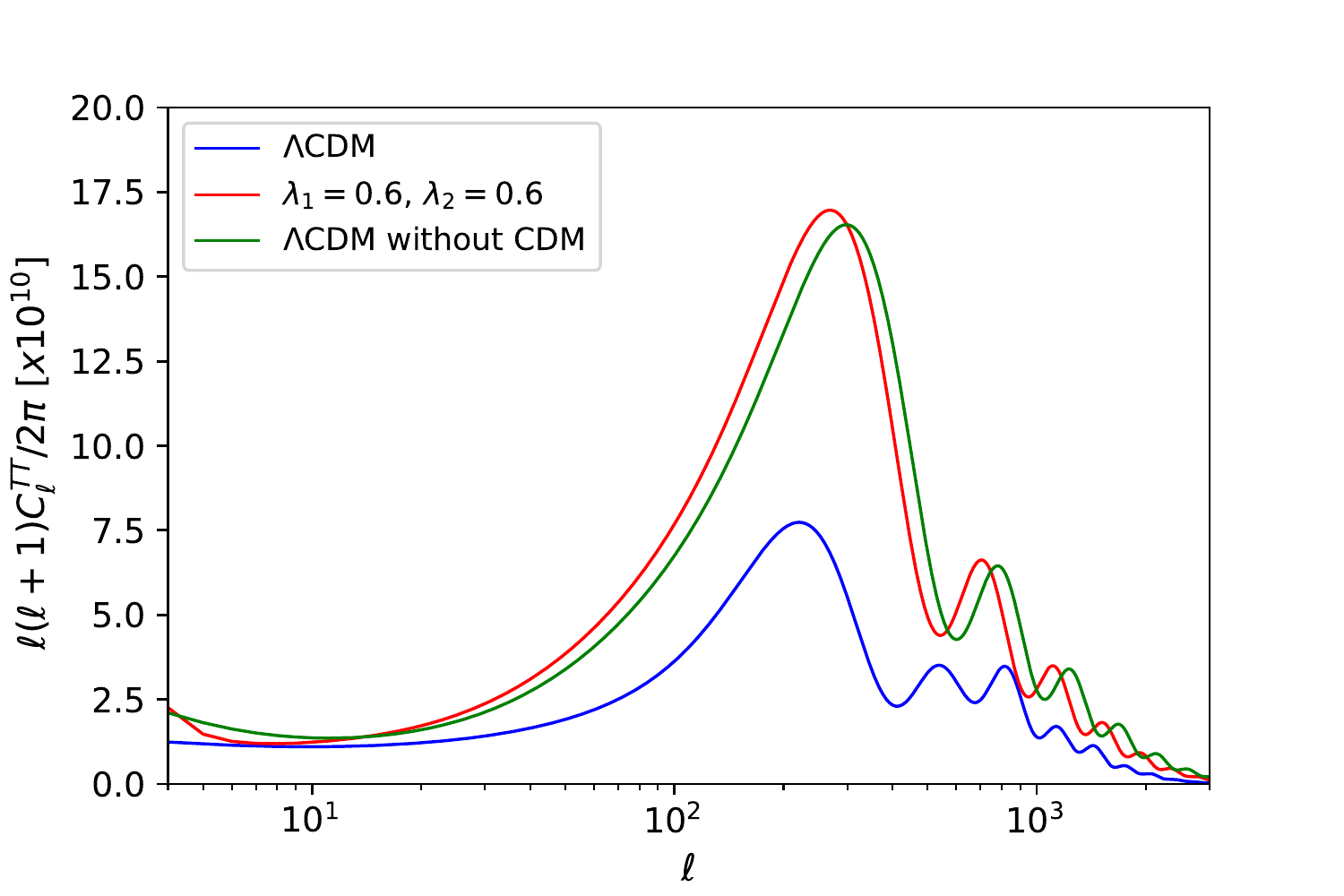}
  \end{subfigure}
\caption{Linear matter power spectrum (left) and dimensionless CMB temperature (right) power spectrum,  for $\lambda_1=\lambda_2=0.6$, $\Lambda$CDM, and $\Lambda$CDM without CDM. The other cosmological parameters were fixed to the \textit{Planck} 2018 best-fit values. }
\label{fig:comparison}
\end{figure*}

A situation where the couplings are large enough to produce the cosmic acceleration leads to a Universe without CDM, as pointed out before. In order to compare the scenarios, we show in Fig. \ref{fig:comparison} the power spectra for the case $\lambda_1=\lambda_2=0.6$ along with $\Lambda$CDM without CDM. We see that the power spectra are very similar to each other, although not identical, because of the different DE equation of state and perturbation equations. In this case, the CMB power spectrum has all peaks increased, when compared to the one for $\Lambda$CDM, due to the absence of CDM, while the third peak is smaller than the second one. On the other hand, the matter power spectrum is reduced mainly on small scales due to the absence of CDM. 

\section{Conclusions}\label{sec:conclusions}

In this paper we investigated an interacting HDE model with the Hubble-scale as the IR cutoff. We assumed that the interaction between CDM and DE is driven by the sum of the energy densities of both species, with constant coupling constants. The evolution of the energy density for both components of the dark sector is the same, leading to  an always constant ratio  $\rho_{\rm dm}/\rho_{\rm de}$ and solving the coincidence problem. However, an analysis of possible values for the couplings that would lead to the cosmic acceleration shows that the corresponding CMB and matter power spectra are very different from the ones in the $\Lambda$CDM model. Hence, this is in disagreement with cosmological observations, indicating that the assumed interacting HDE is not viable to describe the current phase of accelerated expansion of the Universe. 

We  point out that the results presented here are valid for a constant $c^2$. A time varying $c$ changes the DE equation of state and may lead to different conclusions, but it is beyond the scope of the present work.

\begin{acknowledgements}
R.G.L. thanks Matteo Lucca for useful comments and the Group T70 at the Physics Department for the hospitality, while this work was in progress.
\end{acknowledgements}

\bibliographystyle{unsrt}
\bibliography{main}\end{document}